\documentclass[12pt]{article}

\usepackage{amssymb,amsmath, psfrag, epsfig}

\def\u{\vskip  .075 in}

\def\c{\centerline}

\def\u{\vskip  .1 in}

\def\B {\begin{eqnarray*}}

\newcommand{\bel}[1]{\begin{equation}\label{#1}}

\newcommand{\be}{\begin{equation}}
\newcommand{\qe}{\end{equation}}
\newcommand{\ee}{\end{equation}}
\newcommand{\baS}{\begin{eqnarray}}
\newcommand{\ba}{\begin{eqnarray}}
\newcommand{\ea}{\end{eqnarray}}

\newcommand{\bi}{\bibitem}
 
\def\CA{${\rm Ca^{2+}}$ }
\def\CAN{${\rm Ca^{2+}}$}

\def\EN{\end{eqnarray*}}

\begin{document}
\title{Biophysical properties and computational modeling of calcium spikes in serotonergic neurons
of the dorsal raphe nucleus\\
\   \\
{\normalsize
 Henry C. Tuckwell$^{\dagger}$\\   \
$^1$ Max Planck Institute for Mathematics in the Sciences\\
Inselstr. 22, 04103 Leipzig, Germany\\
\     \\
$^{\dagger}$ {\it Correspondence}: tuckwell@mis.mpg.de}  }

\maketitle

\begin{abstract}  
Serotonergic neurons of the dorsal  raphe nuclei,
 with their extensive innervation of nearly the whole brain
 have important  modulatory effects on many cognitive and
 physiological processes. They play important roles in
clinical depression and other psychiatric disorders. 
  In order to quantify the effects of serotonergic transmission
on target cells it is desirable to construct computational models and to this end
 these it is necessary to have details of the biophysical and spike properties of 
the serotonergic neurons. 
Here several basic properties are reviewed with data from several
studies since the 1960s to the present. The quantities included
are  input resistance,
resting membrane potential, membrane time constant, firing rate, spike duration,
spike and afterhyperpolarization (AHP) amplitude, spike
threshold, cell capacitance,   soma and somadendritic areas. 
The action potentials of these cells are normally triggered by 
 a combination of sodium and calcium
currents which may result in autonomous pacemaker activity.
 We here analyse the mechanisms of high-threshold calcium spikes 
which have been demonstrated in these cells the presence of TTX. 
The  parameters for calcium dynamics required to give calcium spikes are quite different from those for regular spiking which suggests the involvement of
restricted parts of the soma-dendritic surface as has been found, 
for example, in hippocampal neurons. 
\end{abstract}
\noindent  Abbreviations: DRN SE, dorsal raphe nucleus, serotonergic.

\noindent Keywords: Serotonin; dorsal raphe nucleus; computational model; spikes.

\section{Introduction}
The spiking activity of neurons in the dorsal raphe nucleus has been studied 
experimentally for many years (Aghajanian and Haigler, 1974; Baraban and
 Aghajanian, 1980) with particular
emphasis on pacemaker-like firing.
Serotonergic neurons in the dorsal and other raphe nuclei,
innervate many brain regions and influence many cognitive, emotional
and physiological processes.  They have been implicated 
in the  pathophysiology of major depressive and other
stress-related psychiatric disorders.
The therapeutical effects of selective
serotonin re-uptake inhibitors indicates the key role of 
serotonin in depression. 

The involvement of serotonergic neurons arises through a variety of physiological,
neurophysiological and endocrine processes. 
Included in the targets of serotonergic neurons of the DRN are the hippocampus, amygdala, locus coeruleus,
prefrontal cortex (PFC) and the paraventricular nucleus of the hypothalamus. 
Serotonergic input from the raphe into
the hippocampus is important for regulating hippocampal
neurogenesis (Balu and Lucki, 2009).  Bambico et al. (2009) found changes
 in  spontaneous 5-HT neuron single-spike
firing activity after chronic uncontrollable stress, and 
changes in the number of spontaneously-active 5-HT
neurons. 

There are many excellent reviews of the properties of these 
cells including, in chronological order, Jacobs and Fornal (1995),
Azmitia and Whitaker-Azmitia (1995), Aghajanian and Sanders-Bush (2002),
 Lanfumey et al. (2005), 
Lowry et al. (2008), Calizo et al. (2011), Del Cid-Pellitero and Garz\'on (2011),
Hale and Lowry (2011), Hayes  and Greenshaw (2011),  Sakai (2011),
Maximino (2012) and Savli et al. (2012).  It is manifest that the spiking activity of the DRN SE neurons is
of prime importance in maintaining serotonin levels throughout the
brain. In this article our focus
is on the a quantitative modeling of this activity which is considered essential in
order to understand the factors which control serotonin release and 
subsequent effects on target cells.

\section{Biophysical and electrophysiological properties of DRN SE neurons}
In this section we tabulate findings on many of the
biophysical, anatomical and electrophysiological properties
of DRN SE neurons, as these properties are important
in the construction and verification of computational models.

\subsection{Input resistance}

\begin{center}
\begin{table}[!h]
    \caption{Input resistance}
\smallskip
\begin{center}
\begin{tabular}{lll}
  \hline
     {\bf Location }   & Input resistance (M$\Omega$) &  Reference (1st au), remarks \\
  \hline
   DRN, rat   & Range 30-76 & Aghajanian (1982) \\
 DRN, rat    &   190 $\pm$ 13   & Crunelli (1983) \\
    
 DRN, rat    &  Ave, 94; range 40-230  & Vandermaelen (1983) \\
DRN rat & 150-400 &   Aghajanian (1984), slice, CH \\
     
       DRN, rat      &  Range 200-400   & Aghajanian (1985) \\
     
       DRN, rat     &  Ave, 143, Range 60-300   &  Segal (1985) \\
     
      DRN, rat     &  Ave, 143, Range 150-400  & Burlhis (1987) \\

    NRM, rat     &   186  $\pm$ 9 & Pan (1990) \\
      
 NRM, g pig   &  255  $\pm$ 50 & Pan (1990) \\
      
 DRN, rat   & 74 $\pm$ 32 & H\'aj\'os (1996), non bursting \\
   
 DRN, rat   & 86 $\pm$ 33 & H\'aj\'os (1996), bursting \\
    
         CRN, rat   &  1600  $\pm$ 200   &  Bayliss  (1997)\\

   DRN, rat   &  175 $\pm$ 23 &  Li (2001) \\
      
    DRN, rat   &  230.1  $\pm$ 8.8   &  Liu (2002)\\
     
         DRN, rat   &  241.5 $\pm$ 29.2   &  Kirby et al., (2003)\\
    
         DRN, MRN rat   &  637 $\pm$ 32, 688 $\pm$ 35  & Beck (2004)  \\
%

DRN, rat & 252  $\pm$ 20 &  Scuve\'e-Moreau (2004) \\
  
 DRN, rat   &  234  $\pm$ 2.5   &  Liu (2005)\\
     
      DRN, mouse    & 427.7 to 580.9  &  Macri (2006),  WT and KO, 3 types \\

       vmDRN, mouse    & 432.4  $\pm$ 34.9  &  Crawford (2010)\\

 lwDRN, mouse    & 592.0  $\pm$ 70.2  &  Crawford (2010)\\
    
 DRN, rat     & 223.8  $\pm$ 5.4  &  Ogaya (2011)\\
 vmDRN, rat     & 600  &  Calizo (2011)\\
 lwDRN, rat     & 500 &  Calizo (2011)\\
 dmDRN, rat     & 550  &  Calizo (2011)\\
 MRN, rat     & 750 &  Calizo (2011)\\
      \hline
\end{tabular}
\end{center}
\end{table}
\end{center}

\newpage
\subsection{Resting membrane potential}

\begin{center}
\begin{table}[h]
    \caption{Resting membrane potential}
\smallskip
\begin{center}
\begin{tabular}{lll}
  \hline
     {\bf Location }   & $V_R$ in mV &  Reference (1st au), remarks \\
      \hline
 DRN, rat     &  -57.7   $\pm$ 1.1  &  Crunelli (1983) \\

       DRN, rat     &  -59   $\pm$ 2.9  &  Segal (1985) \\

     DRN, rat     &  -65  &  Freedman (1987), 1 cell \\
 
         DRN, rat   &  -60   &  Williams (1988) \\
  
   NRM, rat   &  -75 to -50  &  Pan (1990) \\
  
  NRM, g pig   &  -70 to -52  &  Pan (1990) \\
   
         DRN, rat   &  Approx -60 & Penington (1991) dissoc. cell \\

         DRN, rat   &  -65  & Pan (1994), 1 cell \\
      
        CRN, rat   & -49.7 $\pm$ 1.0& Bayliss et al. (1997) \\
       
    DRN, rat   & -68 $\pm$ 13 &  Li (2001) \\
 
   DRN, rat   & -62.4  $\pm$ 1.4 & Liu (2002) \\

      DRN, rat   &  -67.8 $\pm$ 1.4  & Kirby (2003) \\

         DRN, rat      &  63.0 $\pm$ 1.7  &  Beck (2004) \\
    
   MRN, rat      &  66.0  $\pm$ 1.6  &  Beck (2004) \\
    
  DRN, mouse   & -55.3 to -54.6   &  Macri (2006), WT and KO, 3 types \\
  
  vmDRN, mouse   & -66.2  $\pm$ 2.1   &  Crawford (2010) \\

       lwDRN, mouse   & -63.3   $\pm$  1.9   &  Crawford (2010) \\
 vmDRN, rat     & -64.5   &  Calizo (2011)\\
 lwDRN, rat     &  -70 &  Calizo (2011)\\
 dmDRN, rat     & -75  &  Calizo (2011)\\
 MRN, rat     & -64.5  &  Calizo (2011)\\
%
      \hline
\end{tabular}
\end{center}
\end{table}
\end{center}

\newpage 

\subsection{Membrane time constant}

\begin{center}
\begin{table}[!hb]
    \caption{Membrane time constant}
\smallskip
\begin{center}
\begin{tabular}{lll}
  \hline
     {Location }   &  $\tau$ in ms &  Reference (1st au), remarks \\
\hline 
     DRN, rat     &  25.6   &  Crunelli (1983), one cell  \\
       DRN, rat     &   20 to 39  &  Segal (1985) \\
CRN, rat &  42.6 & Bayliss (1997), estimated from $R_{in}$ and C \\
   DRN, rat   & 7.4 $\pm$ 1.4 &  Li (2001) \\ 
DRN, rat & 21.7 & Liu (2002), estimated from Figure 1A \\
        DRN, rat   &  21.4 $\pm$ 4.4  & Kirby (2003) \\
         DRN, rat      &  51 $\pm$ 3.0  &  Beck (2004) \\
   MRN, rat      &  44  $\pm$ 2.2   &  Beck (2004) \\
  DRN, mouse   & 13.9 to 19.5    $\pm$ 0.5  &  Macri (2006), WT and KO, 3 types \\
  vmDRN, mouse &   23.4  $\pm$ 1.9  &  Crawford  (2010) \\
  lwDRN, mouse &   30.6  $\pm$ 2.4 &  Crawford  (2010) \\
 vmDRN, rat     & 49.9  &  Calizo (2011)\\
 lwDRN, rat     & 53.5 &  Calizo (2011)\\
 dmDRN, rat     & 49.3   &  Calizo (2011)\\
 MRN, rat     & 43.8 &  Calizo (2011)\\
      \hline
\end{tabular}
\end{center}
\end{table}
\end{center}

\newpage

\subsection{Firing rate}

\begin{center}
\begin{table}[!hb]
    \caption{Firing rates. CH=Chloral hydrate. PE=phenylephrine.}
\smallskip
\begin{center}
\begin{tabular}{lll}
  \hline
     {Location }   &  Firing rate (Hz)  &  Reference (1st au),  remarks \\
  \hline
DRN rat  & 0.33-0.67 &  Aghajanian (1968), in vivo, CH \\
DRN rat & 1-2 & Sheard (1972), CH, large reduction with amitriptyline\\
DRN rat & 0.5-2 & Aghajanian (1974), inhibited by LSD  \\
DRN rat &  1.3 (mean, 46 units) & Mosko (1974) CH \\
&  & 0.68-1 (10), 1.01-2 (14), $> 2.01$ (10)  \\
DRN rat  & 0.5-2.5    & Gallager (1976), CH  \\
DRN cat & 0.5-5     &  McGinty (1976), in vivo, quiet waking, Exc Histos.  \\
& & W 2.4 $\pm 1.3$, SWS 1.3  $\pm 0.8 $, REM 0.2  $\pm 0.3$ \\
DRN rat & 0.8-2.5  & Mosko (1976),  slice. Histograms of ISIs \\
 DRN rat   &   0.25-2& Aghajanian (1978), CH, suppressed by sciatic n. stimulation \\
DRN cat  & 2.82   $\pm 0.17 $  &   Trulson (1979), freely moving  \\
DRN rat   & 1.7 (1 cell) & Aghajanian (1982) CH, ceased spiking with LSD \\
DRN rat & 1.1( 1 cell) & Crunelli (1983), slice \\
DRN rat & 1-3.5  & Vandermaelen (1983), slice, PE applied \\
 DRN rat & 0.5-5, mean 2 &  Chu (1984), most cells inhibited by ethanol \\
DRN rat & 0.5-3 &   Sprouse (1987), CH \\
DRN rat  & 0.5-1.5 & Park (1987) \\
DRN, MRN rat & 0.1-3 &  H\'aj\'os (1995), in vivo, CH, 270 of 372 non bursting \\
DRN rat & 1-4 &  H\'aj\'os (1996), in vivo, regular (non bursting)  \\
DRN rat & 1.2  $\pm 0.1 $ (31)  &  H\'aj\'os, Sharp (1996), in vivo, CH, regular (non bursting)  \\
& 1.0  $\pm 0.1 $ (22)  &   bursting, 49.2   $\pm 6.9 $ \% spikes in bursts  \\
DRN rat & 0.968  $\pm 0.12 $ (23) & Kinney (1997), CH, control. HISTOS\\
 & 0.417  $\pm 0.071 $  &  clomipramine treated neonates, 'depressed' \\
CRN rat &  1.3  $\pm 0.2 $ (24)   & Bayliss (1997), slice\\
DRN cat & 3   &  Jacobs (1999), quiet waking, less during sleep \\
&   &  stressors (acute), no change\\
DRN rat & 1.18   $\pm 0.07 $  & Kirby (2000), in vivo, Halothane\\
  &  &  predominantly inhibited by CRF \\
DRN rat  &  ISI 120 ms & Li (2001), slice, 0.3 nA  \\
  & &  ISI 85 ms  0.4 nA;  ISI 75 ms 0.5 nA  \\
DRN rat & 1.67  $\pm 0.24 $ (24)   & Allers (2003), in vivo, range 0.37-3.0 \\

      \hline
\end{tabular}
\end{center}
\end{table}
\end{center}

\begin{center}
\begin{table}[!ht]
    \caption{Firing rates, continued.  CH=Chloral hydrate.
            CUS=chronic unpredictable stress. \c{ FS=footshock. MBR=midbrain raphe. U=urethane.} }
\smallskip
\begin{center}
\begin{tabular}{lll}
  \hline
     {Location }   &  Firing rate (Hz)  &  Reference (1st au),  remarks \\
  \hline
DRN rat & 0.8   $\pm 0.3 $ (3); 0 (14)   & Kirby (2003), slice, spontaneous \\
&    0.4   $\pm 0.2 $ (5)       & elicited by PE\\
DRN dog &  1.18 $\pm 0.13$  (19)  &  Wu (2004), in vivo, REM off, active waking \\
   & 0.32 $\pm 0.05$ (19)  & REM off, non-REM sleep \\
   &   0.07 $\pm 0.02$ (19)   & REM off, REM sleep \\
  &  1.42  $\pm 0.18$ (16) & REM-reduced, active waking\\
  &  0.81 $\pm 0.20$ (16) & REM-reduced, non-REM sleep \\
  &  0.57 $\pm 0.09$ (16)& REM-reduced, REM sleep)\\
&  & For other cell types see Table 2 in Wu (2004) \\
DRN rat & 1-5 & Waterhouse (2004) freely moving, quiet rest \\
DRN rat  & 1.1 $\pm 0.13$   &  Liu (2005), slice, PE 3 $\mu$M\\
MRN rat &  0.52 $\pm 0.03$  (41)  &  Judge (2006), slice, PE 1 $\mu$M, inhib by 5-HT\\
DRN rat &  1.21 $\pm 0.07$  (42)  &  Judge (2006), slice, PE 1 $\mu$M, inhib by 5-HT\\
MBR rat  & 5.40  $\pm 1.95 $  (10) & Kocsis (2006), in vivo, U, during hippocampal theta\\
 &  4.69  $\pm 1.94 $  (10)  & non hippocampal theta \\
DRN rat & 1.4 $\pm 0.4 $ (10)   & H\'aj\'os (2007),CH,  bursting, range 0.4-4.1 \\
   & ISI 7.7 $\pm 0.4 $ in bursts  &  14.7 \% of spikes in bursts, has HISTOS  \\
DRN rat & 0.1-3.5   &  Bambico (2009), in vivo, CH, control   \\
     & &   ISI $\approx$ Gaussian, HISTOS \\
  &   Decreased 35.4\% & CUS group (101), ISI bimodal, more bursts\\
vmDRN mouse & 1.1  $\pm 0.3 $ (7) &  Crawford (2010), slice, PE 1 $\mu$M \\
lwDRN mouse & 2.2 $\pm 0.4 $ (6)  &   Crawford (2010), slice, PE 1 $\mu$M  \\
DRN rat & 2.1 $\pm 0.2 $ (8) regular  & Schweimer (2010), in vivo, isoflurane, U\\
 &   &  some excited by footshock, some no change \\
    & 1.5   $\pm 0.18  $ (11) bursting & some excited, some no change, some inhibited by FS\\

      \hline
\end{tabular}
\end{center}
\end{table}
\end{center}

\subsection{Spike duration}

There have been many recordings of spiking activity in serotonergic neurons of the
DRN and other raphe nuclei. 
A distinguishing feature of spikes in such neurons is their long duration, 
usually taken to mean with 
magnitude equal or
greater than about 2 ms.  Other cells in the DRN, principally presumed to be inhibitory GABA-ergic neurons,
have characteristically shorter spike durations, around 1 ms (Liu et al, 2002;
Liu et al, 2005). 
 Some early recordings did not give spike durations (Trulson  and Jacobs, 1979; 
Crunelli et al, 1983; Aghajanian 1985; Segal, 1985) and several more recent data on spike durations have been 
obtained for extracellular recordings
(Allers and Sharp, 2003; Waterhouse et al., 2004; Kocsis et al., 2006; Urbain et al. 2006; 
Nakamura et al. 2008; Schweimer and Ungless, 2010; Sakai, 2011). There are also data for
embryonic or neonatal animals (Li and Bayliss, 1998; Moruzzi et al., 2009).   Representative  
results for intracellular recordings are given in Table XX. These 
were either given explicitly or estimated (denoted by an asterisk)
 from figures in the 
manuscripts. The definition of spike duration differs
from author to author and the method of evaluation has also been reported
in the Table if it is available.  Also stated are factors which lead
to a lengthening of the spike duration such as EGTA 
or in some cases TTX which, presumably, leads to purely calcium
spikes. Temperature also influences spike duration as noted 
in the original work of Hodgkin and Huxley (1952). In experiments
where the temperature was not at room temperature (20-25$^\circ$C)
the temperature is included in the Table entry. 

\begin{center}
\begin{table}[!hb]
    \caption{Spike duration, intracellular studies; $^*$ denotes estimated.}
\smallskip
\begin{center}
\begin{tabular}{lll}
  \hline
     {Location }   &  Duration (ms)  &  Reference (1st au),  remarks \\
  \hline
       DRN, rat     &   2.9$^*$  &   Park  (1982), at -50 mV, antidromic  \\

       DRN, rat     &   2  &  Aghajanian (1982), 35-37$^\circ$C\\
DRN rat & $\sim$2 &  Aghajanian (1984), slice, CH\\
   
     NRM, rat \& g pig     &  1.5  &   Pan (1990), at threshold, 37$^\circ$C \\
  
  DRN, rat   &  10$^*$  & Penington (1991), TTX, EGTA 10mM, Dissoc cell \\
     
 DRN, rat   &  3.3$^*$  & Penington (1992), at -50 mV, EGTA 11 mM, Dissoc cell \\
      
 DRN, rat   & 1.25 $\pm$ 0.16  & H\'aj\'os (1996), at half amplitude \\
    
        CRN, rat   & 5.5   & Bayliss  (1997), at half amplitude, EGTA 0.2 mM \\
      
   DRN, rat   & 3.6  $\pm$ 1.2 &  Li (2001), at base,  34-35$^\circ$C \\
    
    DRN, rat   &   0.99   $\pm$ 0.01   & Liu  (2002), at half amplitude \\
 
        DRN, rat   & 2.0     & Kirby  (2003), 35$^\circ$C \\

         DRN, MRN rat      & 2.0,2.1  $\pm$ 0.1 &  Beck  (2004), 32$^\circ$C, EGTA 0.02 mM, ventromedial DR \\
%

DRN, rat      & 3.0  $\pm$ 0.3   &  Marinelli (2004), at threshold, 34$^\circ$C, EGTA 11 mM  \\

DRN dog & 1.64  $\pm$ 0.06  & Wu (2004), cells off during REM \\
DRN dog & 1.44  $\pm$ 0.06  & Wu (2004), cells active when awake \\

  DRN, mouse   & 3.8 to 4.6   &  Macri  (2006), at threshold, EGTA 11 mM\\
     DRN rat &     2.3  $\pm$ 0.3  &    H\'aj\'os (2007), bursting cells, range 1.4-4.0 \\
            vmDRN mouse    & 1.8   $\pm$ 0.1   &  Crawford (2010), at half amplitude, EGTA 0.02 mM \\
  
 vmDRN rat     & 2.1 &  Calizo (2011)\\
 lwDRN, rat     & 2.2  &  Calizo (2011)\\
%
      \hline
\end{tabular}
\end{center}
\end{table}
\end{center}

\newpage 
\subsection{Spike amplitudes and AHP}       
The definition employed for spike amplitude varies. A natural choice is from resting
potential to the peak voltage of the spike, but many authors use the difference
between peak voltage and threshold.  Similarly for the AHP amplitude.
If the definition  has been stated in the article, it is mentioned
in the table. If it has been estimated from a figure then the first definition has been employed. 
Spike amplitude can be quite variable, one factor being dendritic
morphology (Petterson and Einevoll, 2008).

\begin{center}
\begin{table}[h]
    \caption{Spike amplitude and AHP}
\smallskip
\begin{center}
\begin{tabular}{llll}
  \hline
     {Location }   &  Spike amplitude & AHP amplitude  &  Reference (1st au), remarks \\
  \hline
    DRN, rat     &  62 to 80   &  $>6$ & Aghajanian (1982) \\
  DRN, rat     &  70   & 10 & Crunelli (1983) \\
  DRN, rat     &  92    &  16 & Aghajanian (1985) \\
DRN rat  & 87 & 10 & Burlhis (1987)  \\
DRN rat & 85 to 95 & - & Freedman (1987) \\
NRM rat & 83 & 14.2 & Pan (1990) \\
CRN rat & 74 & 13 & Bayliss (1997) \\
DRN rat & 73 & 12 & Liu (2002) \\
DRN rat & 61.4 & 15.9 & Kirby (2003) \\
DRN rat & 69 & 16 & Beck (2004), from threshold\\
MRN rat & 72 & 20 & Beck (2004), from threshold \\
DRN mouse & 66.4 to 71.7 & 19.9 to 21.4 & Macri (2006)\\
vmDRN mouse & 56.6 & 29.4 & Crawford (2010), from threshold\\
lwDRN mouse & 62.2 & 33.7 & Crawford (2010), from threshold\\
 vmDRN, rat     & 71.2 & 14.8   &  Calizo (2011)\\
 lwDRN, rat     & 90 & 15.7&  Calizo (2011)\\
 dmDRN, rat     & 82.5 & 12.9   &  Calizo (2011)\\
 MRN, rat     &  75 & 21.3  &  Calizo (2011)\\
         \hline
\end{tabular}
\end{center}
\end{table}
\end{center}

\subsection{Threshold}
Voltage threshold for action potentials is that
membrane potential at which the sharp depolarizing
phase of a spike occurs.  Such a threshold is usually
around 10 mV above resting potential. 
Data on spike thresholds in serotonergic neurons of the DRN
are sparse although there are some published results
which give such quantities as around -20 mV, which makes
their interpretation difficult, as resting potentials
themselves are around -60 mV.  The few available
results which are compatible with the above definition
are given in the following table.

\begin{center}
\begin{table}[h]
    \caption{Thresholds}
\smallskip
\begin{center}
\begin{tabular}{lll}
  \hline
 Source  & Threshold (mV)  & Remarks  \\
  \hline
Burlhis (1987), rat DRN  & -60  &  Depolarizing pre-spike potential\\
 & -40 &  High threshold spike (\CAN) \\
 H\'aj\'os (1996), rat DRN & -44 $\pm 3$ &  Bursting and non-bursting cells\\
Kirby (1997) rat DRN & -54.9  $\pm 0.5$ & Resting potential -67.8   $\pm 1.4$ \\
Macri (2006), mouse DRN, type 1 & -43.6   $\pm 0.1$  & Resting potential  -54.6   $\pm 0.5$ \\

         \hline
\end{tabular}
\end{center}  
\end{table}
\end{center}

\subsection{Estimates of capacitance}

Estimates of cell capacitance can be made from
measured currents in response to a given change of voltage 
or deduced from either anatomical (surface area) or electrophysiological
measurements of input resistance and time constant. 

\subsubsection{Direct measurement}
There have been very few reports of capacitances of DRN cells.
For a dissociated rat  serotonergic neuron, Penington and Fox (1995) gave a 
 capacitance of 0.02 nF. Bayliss et al. (1997) reported a value of
26.6 $\pm$ 0.09 pF for 32 cells in rat caudal raphe nucleus,
but these cells had a very high average input resistance
compared with other sets of results (see Table 1). For mouse
DRN Macri et al. (2006) found capacitances between 30 and 55 pF.
  For 38 cells in rat DRN, Marinelli et al. (2004) obtained a capacitance
of 39 $\pm$ 3 pF.  Considering the hyperpolarizing 
response to 0.20 nA current injection in a rat DRN cell given
in Figure 1 of Liu et al. (2002), a crude estimate of
the capacitance is 280 pF, which is greater than any of the
values estimated from time constant and input resistance
listed in Table 7.

\subsubsection{Estimates from surface area}
  Visualizations  of complete serotonergic (SE) neurons are not
feasible
 because
of their extensive axonal arborizations, but  a useful schematic
is shown in Figure 1 taken from Maeda et al. (1989). 

      \begin{figure}[!t]
\begin{center}
\centerline\leavevmode\epsfig{file=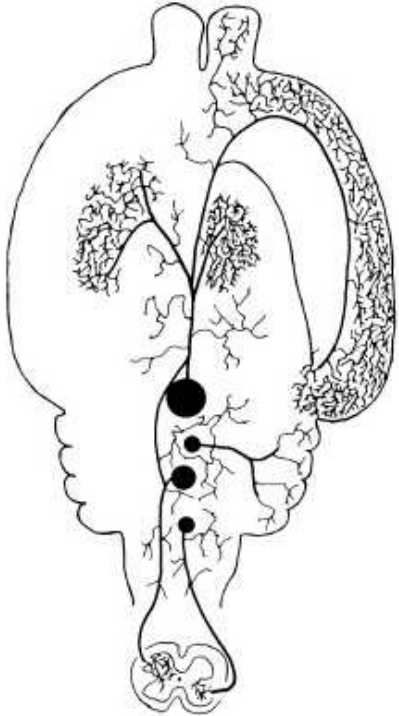,width=2.75in}
\end{center}
\caption{Illustration of the extensive innervation of brain ans spinal cord regions by serotonergic
neurons of raphe nuclei. From Maeda et al. (1989).} 
\label{fig:wedge}
\end{figure}

There have been numerous morphological studies of
the somadendritic parts of these cells (for example, Pfister and Danner, 1980;
Diaz-Cintra et al., 1981;   Descarries et al., 1982;
Park et al., 1982; Imai et al., 1986; Park, 1987;  Li et al., 2001; 
 Allers and Sharp, 2003; 
Haj\'os et al., 2007;   Calizo et al., 2011).  
The term ``soma area'' is used in various  ways. In one early study of rat DRN SE neurons, Descarries et al. (1982) 
gave a seemingly small cell body area
 of 251 $\mu^2$, but this was cross-sectional area, which is made clear by the fact that the 
volume was 2999 $\mu^3$ so that, assuming, as did Descarries et al., a spherical 
shape gives a diameter of 17.9 $\mu$.  Such a geometry gives a soma surface area of
1006  $\mu^2$.

  A clear image of rat DRN SE neuronal cell bodies, dendrites
and proximal axonal branches appeared in Allers and Sharp (2003) and from this study
the soma may be approximately considered a prolate ellipsoid.  
The surface area of such an ellipsoid with semi-axes of $a < b$ is
\be A= 2\pi a^2\bigg(1 + \frac{b}{a\rho} arcsin \rho \bigg) \ee
where 
\be  \rho=\sqrt{1 - \frac{a}{b}^2} \ee
With the estimates of axes $b=16$  $\mu$ and $a=8$ $\mu$ this gives a soma area
 of 1375 $\mu^2$. 
From the same source the dendritic surface
area is estimated approximately  as 2500 $\mu^2$ 
which gives a total SD area of 3875 $\mu^2$.

Detailed morphological data on three kinds of cells in male rats of  ages 30, 90 and 220 days 
were presented by Diaz-Cintra et al. (1981).
Considering the results for the fusiform type (those with the largest cell bodies) in 30 day old rats
leads to the area estimates in the following table TZ. Here two sets of results are
given based on mean values and mean + 1 standard deviation. 
The soma areas are calculated assuming prolate spheroids, as in the above formula. 
The dendritic areas are divided into primary and secondary as in Diaz-Cintra et al. (1981),
for which length and diameter at midpoint were given. Spines have not been taken into account and
the calculated areas are approximate only but indicative.  In Calizo et al. (2011) total
dendritic lengths were given but not diameters.  In the estimate of dendritic area
the same decomposition given by Diaz-Cintra et al. (1981) in terms of primary
and secondary dendrites as well as diameters were employed. 

It can be seen that the estimated soma area, the total dendritic area
and the total SD area, using the mean properties
of  Diaz-Cintra et al. (1981) are not far from the values
estimated for the Allers and  Sharp (2003) neuron. Further,
using the mean + 1 standard deviation for the properties
of the Diaz-Cintra et al. (1981), leads to estimates which are
very similar to those for the vmDR data of Calizo et al. (2011).
The range of SD areas so obtained is from about 3900 $\mu^2$ to 
11600 $\mu^2$.  The fraction of total surface area which is somatic
has values 0.27, 0.19, 0.35 and 0,22 with an average value
of 0.26.
Clearly if one used figures  less than the means of
Diaz-Cintra et al. (1981),  the SD area would be perhaps as small as
2500 $\mu^2$.  In summary, the SD area of the serotonergic cells
in the DRN has an approximate estimated mean value of 3900 $\mu^2$,
with a lower limit around 2500 $\mu^2$ and an upper limit of around 11600  $\mu^2$.

\begin{center}
\begin{table}[h]
    \caption{Estimates of soma-dendritic area of DRN serotonergic neurons in $\mu^2$}
\smallskip
\begin{center}
\begin{tabular}{lllll}
  \hline
Quantity & Means & Means + 1SD & Allers (2003) & Calizo (2011)  \\
 &  Diaz (1981) &  Diaz (1981)  & & vmDR   \\
  \hline
Soma area  &  1073  & 2197 & 1375  & 2400  \\
Primary dendtites     &  1888  &  6148 & & 6083 \\
Secondary dendrites &  953   &  3300 &  &   2675  \\
SD total area  &  3914 & 11645 & 3875 & 11158 \\
Fraction soma & 0.27 & 0.19 & 0.35 & 0.22\\
Fraction dendrites & 0.73 & 0.81 & 0.65 & 0.78\\
         \hline
\end{tabular}
\end{center}
\end{table}
\end{center}

Using the standard figure of $1\mu F$ per sq cm,  the estimated area
from the Allers and Sharp (2003) study would give a capacitance of about 0.039 {\it nF}.
The same value is obtained from the average data, as explained above,
 in Diaz-Cintra et al. (1981). Using the smallest estimated area, the
capacitance would be about  0.025 nF (whole cell) and using the largest area,
 C would be about 0.116 nF.

\subsubsection{Capacitance from time constant and input resistance}
One may use  time constants $\tau$
and input resistances $R_i$ as given in Tables tz and fg to obtain an
approximate estimate of $C$  from the formula $\tau=CR_i$. This can sometimes
yield accurate results (Reyes et al., 1994).  Using this approach gives the estimates in the
following table.

\begin{center}
\begin{table}[!ht]
    \caption{Capacitance estimated from $\tau$ and $R_{in}$}
\smallskip
\begin{center}
\begin{tabular}{llll}
  \hline
     {Source (1st au) }   &   C in pF &  Source (1st au) & C in pF \\
      \hline
Crunelli, DRN rat (1983)  &  135  &  Macri, DRN mouse (2006)  & 34 \\
     Segal, DRN rat (1985) & 210 & Crawford, vmDRN mouse (2010) & 53  \\
Li, DRN rat (2001)  & 40 & Crawford, lwDRN mouse (2010)    & 52  \\
Liu, DRN rat (2002)  & 94 & Calizo, vmDRN rat (2011) & 83  \\
Kirby, DRN rat  (2003)   & 87 & Calizo, lwDRN rat (2011)  &   108\\
Beck, DRN rat  (2004)   & 80 &  Calizo, dmDRN rat (2011) &  89 \\
Beck, MRN rat  (2004)   & 64 &  Calizo, MRN rat (2011) & 59  \\

      \hline
\end{tabular}
\end{center}
\end{table}
\end{center}

 Not including the possible outlier of 206 pF, 
 the range of these estimates in rat DRN is from 40 pF to 135 pF with an average
of 89.5 pF.

\subsubsection{Summary of estimates of capacitance}
  Table FF summarizes the various results obtained by the three
different methods (Measurement, Deduction from area, $\tau/ R_i$). Most of these
estimates
are heuristically obtained and will doubtless be superseded
 when more
detailed measurements are made.

\begin{center}
\begin{table}[!ht]
    \caption{Summary of estimates of capacitance}
\smallskip
\begin{center}
\begin{tabular}{llll}
  \hline
   Quantity  &  Measurement &  Estimate from areas & From $\tau/ R_i$  \\
      \hline
Mean & 39 pF  & 55 pF   &  89.5 pF \\
Most likely & &   39 pF  &  \\
Range & 39 $\pm 3$ pF & 25-116 pF & 40-135 pF\\
      \hline
\end{tabular}
\end{center}
\end{table}
\end{center}

To construct a computational model of a "typical" non-dissociated
 DRN serotonergic neuron,
one might employ a capacitance of 40 pF, a total area of 4000 $\mu^2$, 
and a soma area of 1000  $\mu^2$.  For modeling a larger cell, these quantities
could be chosen as 90 pF, 9000  $\mu^2$ and 1800  $\mu^2$, respectively.

\section{Computational modeling of calcium spikes in TTX}
  In their pioneering study of the electrophysiology and ionic basis of pacemaking
of presumed serotonergic
neurons of the dorsal raphe nucleus, Burlhis and Aghajanian (1987)
performed several    ``anodal break'' experiments. In one of these TTX was
applied to block fast sodium currents and under
 the application of a steady depolarizing current of 0.15 nA there were obtained spikes, called
high-threshold calcium spikes. For the cell in which such spikes were
depicted (see Figure 3D of
Burlhis and Aghajanian, 1987), the interspike interval (ISI) was about 300 ms
and the spike amplitude was about 75 mV. These data may be compared
with the usual spike properties of an ISI of around 1000 ms and a spike amplitude
around 90 mV. High and low threshold calcium spikes in these neurons  with TTX were also demonstrated by Penington et al. (1991). 

In a forthcoming article (Tuckwell and Penington, 2012) a computational
model for spiking in DRN SE neurons is developed which contains
7 voltage-dependent ion currents, $I_i$,  a leak current, $I_{Leak}$, a  calcium-activated potassium current $I_{SK}$ and a calcium- and voltage-activated
potassium current  $I_{BK}$, BK and SK denoting big and small channel
conductances.  The model is completed with calcium dynamics,
in which there are sources due to the calcium currents as well as buffering and pumping.

The differential equation for the membrane potential $V$ can be written 
\be  C\frac{dV}{dt}=-\bigg[ \sum_iI_i + I_{Leak} +  I_{SK} +  I_{BK} +\mu \bigg] , V(0)=V_0, \ee 
where C is the capacitance, $V_0$ is the initial
value of $V$, taken to be the resting membrane potential, $V_R$. 
and 
an applied current $\mu$  is added.
Depolarizing currents are negative, all voltages are in mV, and $t$ is in ms so that if $I$ is in 
nA, then the membrane capacitance is in nF. 
For the  $I_i$,  Hodgkin-Huxley type  (1952) formulations are used 
generically and generally, each component current being taken as a product of activation 
and inactivation variables,  a maximal conductance, $g_{i,max}$, 
and a driving 
force which is $V-V_i$ where $V_i$ is usually at or near the Nernst equilibrium
potential. 

For noninactivating currents there is  an activation variable $m$ raised to a certain power $p \geq 1$,
not necessarily an integer, 
so that 
\be  I_i=g_{i,max}m^p(V-V_i).  \ee
If the current inactivates, then the current contains an inactivation variable $h$ which is usually 
 raised to the power 1 so 
\be I_i=g_{i, max}m^ph(V-V_i). \ee
For the calcium dependent currents the activation variables depend
 on, or also on,  calcium ion concentration. 

For each $I_i$, activation and inactivation variables are determined by differential equations
\be \frac{dm}{dt} = \frac{m_{\infty} - m}{\tau_m} \ee
\be \frac{dh}{dt}=\frac{h_{\infty} - h}{\tau_h} \ee
where $m_{\infty}$ and $h_{\infty}$ are steady state values which depend on voltage.
The quantities $\tau_m$ and $\tau_h$ are time constants which may also
depend on  voltage and/or calcium concentration.
For an account  of the dynamics of many types of current Destexhe and Sejnowski (2001).

      \begin{figure}[!h]
\begin{center}
\centerline\leavevmode\epsfig{file=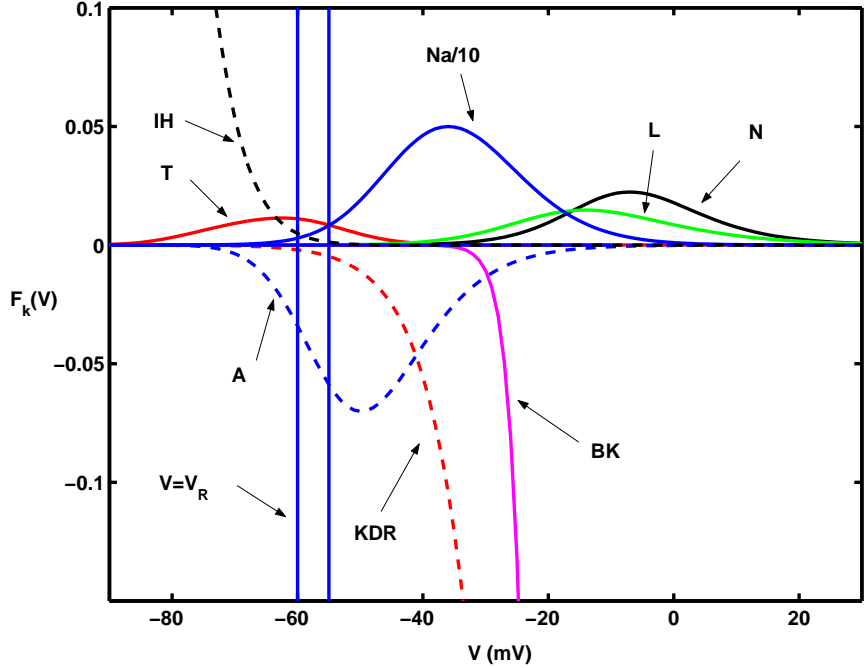,width=4.5 in}
\end{center}
\caption{Contributions to the source function $F$ from various component
currents in a case which leads to spontaneous spiking with sodium and calcium
as the inward currents. Key: A = transient potassium, Na = fast sodium (divided by 10), T = low threshold T-type calcium, L and N, high threshold L and N-type calcium,
KDR = delayed rectifier porassium, BK = big conductance calcium-activated
potassium, here approximated with a voltage-dependent form as in Tabak et al. (2011). The vertical lines mark the resting potential $V_R$ and $V_R + 5$.} 
\label{fig:wedge}
\end{figure}

      \begin{figure}[!h]
\begin{center}
\centerline\leavevmode\epsfig{file=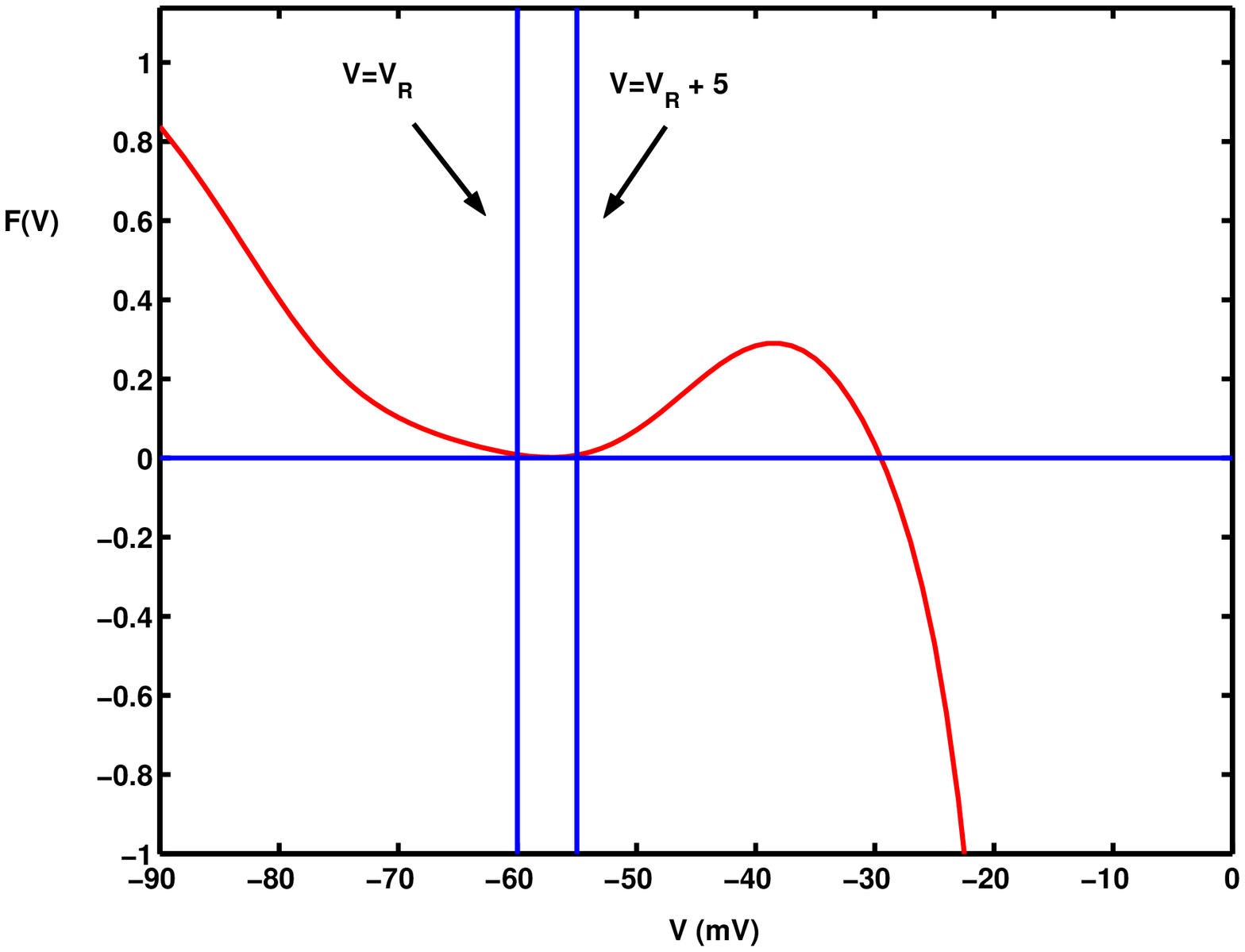,width=4.5 in}
\end{center}
\caption{The sum $F(V)$ of the terms in the previous figure, obtained with
 no applied current so $\mu=0$. Vertical lines as in Figure 2.} 
\label{fig:wedge}
\end{figure}

The differential equation describing the evolution of the
internal 
calcium ion concentration $Ca_i$  is similar to 
that employed by  Rybak et al. (1997), 
\begin{equation} 
\frac{dCa_i} {dt}= -CSF(I_L +I_N).\frac{1-PB(t)}{2Fv} - K_s. \frac{Ca_i}{Ca_i + K_m}  \end{equation}
where the fraction of calcium which is bound is 
\begin{equation} PB(t)= \frac{B_{tot}}{Ca_i + B_{tot} + K_d},  \end{equation}
$K_d$ being the dissociation constant. 
In (8), $I_L$ and $I_N$ are (relatively) high-threshold L-type and N-type
 calcium currents,
CSF is a calcium source factor, F is Faraday's constant,  v is the volume of the
internal shell housing internal calcium ions, $K_s$ is the pump strength and
$K_m$ is the pump half-activation concentration. Concentrations are
in mM.  

      \begin{figure}[!h]
\begin{center}
\centerline\leavevmode\epsfig{file=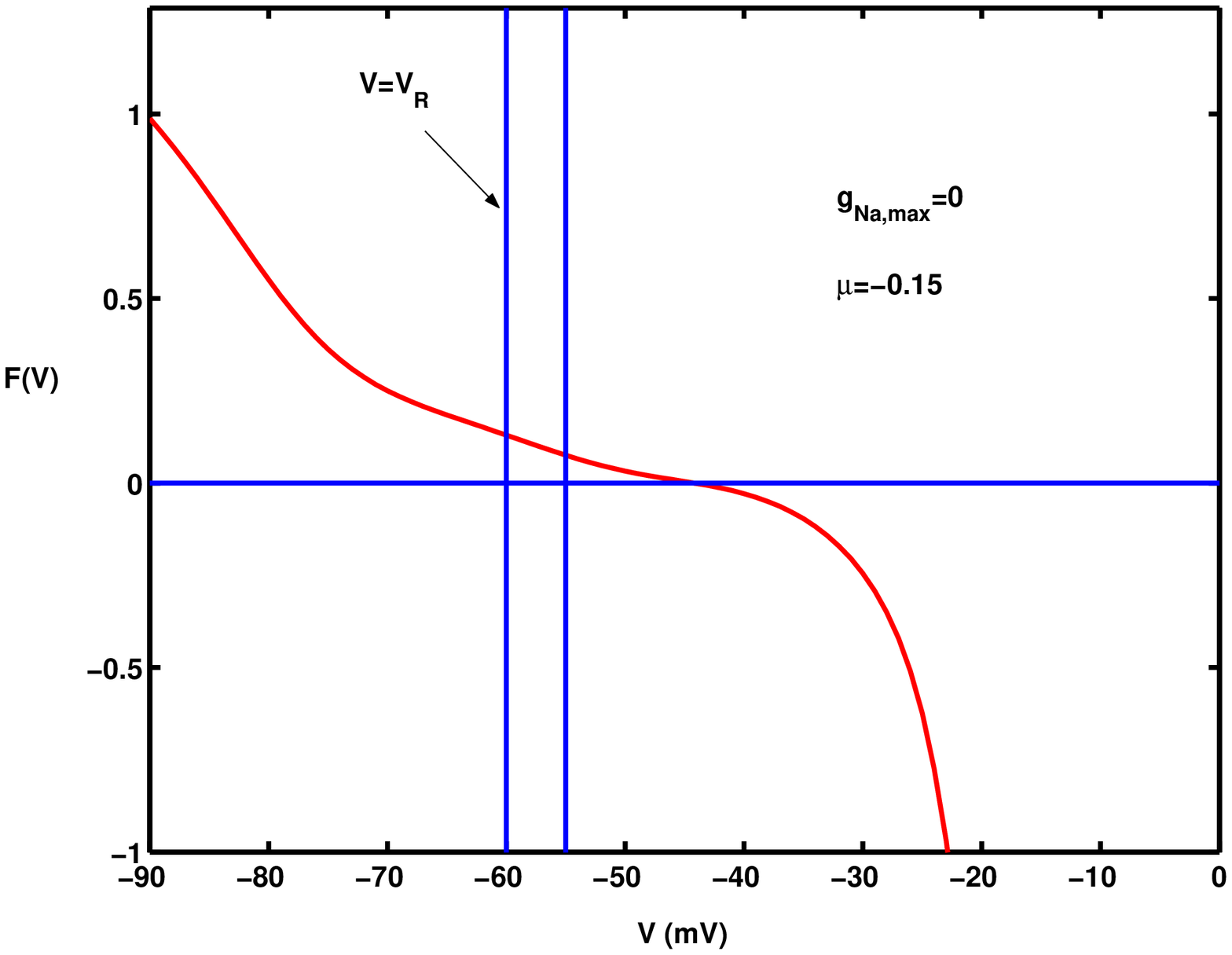,width=4.5 in}
\end{center}
\caption{A plot of $F(V)$ with all terms as in Figure 2, but that now
there is no sodium conductance and there is an applied depolarizing
current of magnitude 0.15 nA.} 
\label{fig:wedge}
\end{figure}

In order to ascertain approximately if a given set of parameters will
result in spiking, define for each voltage-dependent channel type, $k$,
\be F_k(V) = g_{k,max}m_{k, \infty}(V)^{p_k}h_{k, \infty}^{q_k}(V) (V-V_k) \ee
 being  the contribution of the $k$-th current at steady state. 
Here $p_k$ and $q_k$ are the powers to which the activation and inactivation
variables are raised. In all cases encountered,  $q_k$ is either 0 (no inactivation) 
or 1.  For the full model, including fast sodium current, with no applied
current, spiking can be spontaneous and in such a case the functions $F_k$ 
are shown plotted against $V$ in Figure 2. 
The total source function 
\be F(V) = \sum_k F_k(V) \ee
is shown in Figure 3. It can be seen that $F>0$ for $V$ in an interval
containing the resting potential, which makes the deriviative of $V$ positive
at rest, so that no stimulus is required to make the cell fire.  The curve $F$
is similar to the cubic in the Fitzhugh-Nagumo model. Although $F>0$ is
an approximate necessary condition for firing, it is not sufficient as will
be seen shortly. 

      \begin{figure}[!h]
\begin{center}
\centerline\leavevmode\epsfig{file=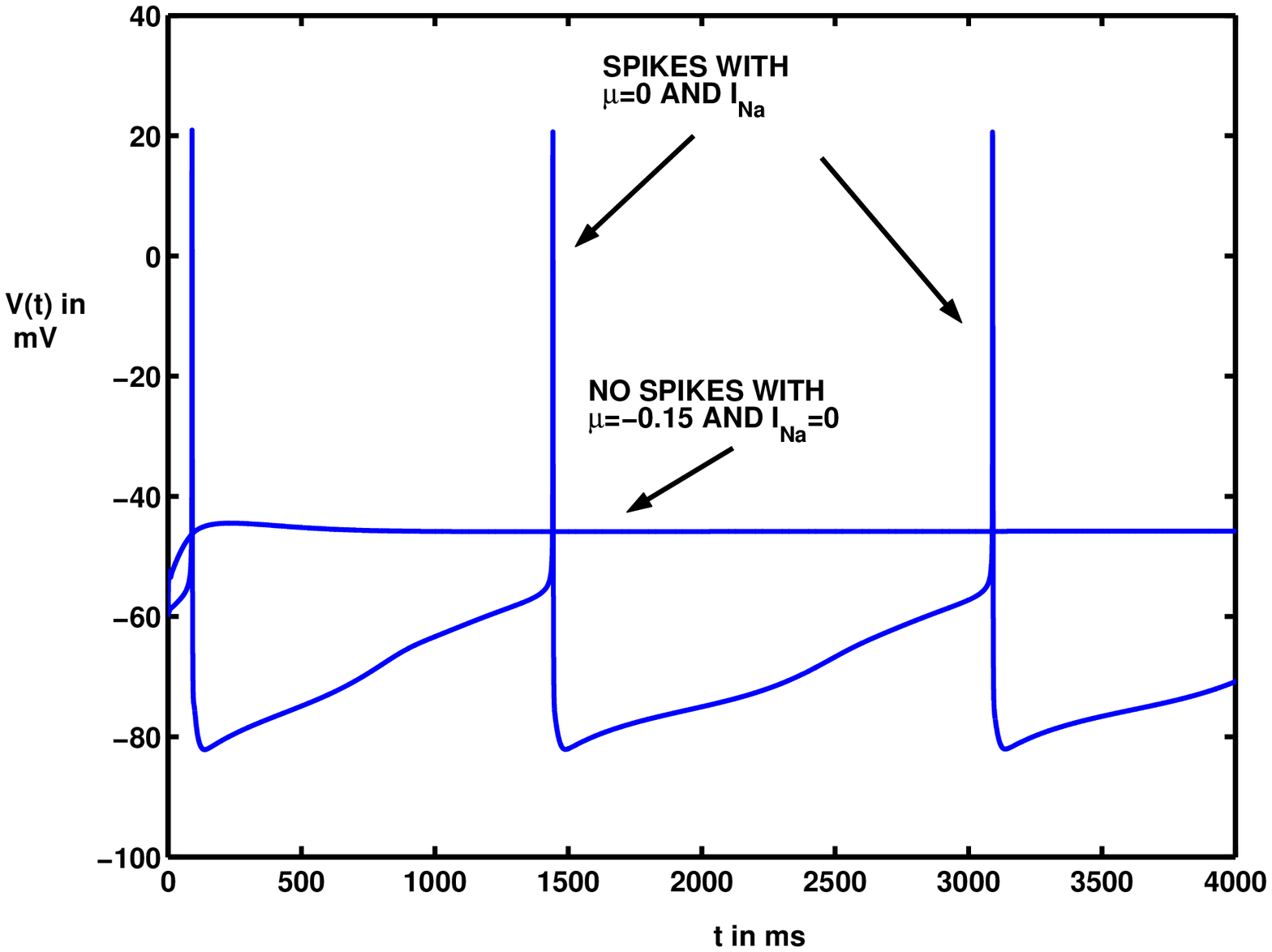,width=4.5 in}
\end{center}
\caption{Membrane potential versus time for the two cases whose $F(V)$s are
plotted in Figures 3 and 4. Whereas spontaneous activity ensues with
sodium current, without the sodium current there are no spikes
even with an applied depolarizing current of 0.15 nA.} 
\label{fig:wedge}
\end{figure}

When there is no fast sodium current, as is the case when 
TTX is applied, and when $\mu=-0.15$, as in the experiment described
above by Burlhis and Aghajanian (1987), the resulting function $F(V)$ is as
depicted in  Figure 4. Here all the parameters of all the other
component currents are the same.

 In Figure 5 are shown plots of the contrasting voltage responses versus time 
for the two cases of source functions depicted in Figures 3 and 4. 
For the case with sodium current and no applied
current, the cell spontaneously fires action potentials in a pacemaker-like fashion
with an ISI of about 1600 ms. In the case of no sodium current and
an applied depolarizing current, it is still the case that $F>0$ in a large interval containing the rest point,  but there are no spikes. There is a depolarization
 to -44.5 mV from the resting value of -60 mV, followed by a 
very slow decline to -45.8 mV at 4000 ms. Clearly, in order to obtain
spikes (calcium spikes) for the model in the absence of fast sodium current, changes in
other parameters is necessary.

      \begin{figure}[!h]
\begin{center}
\centerline\leavevmode\epsfig{file=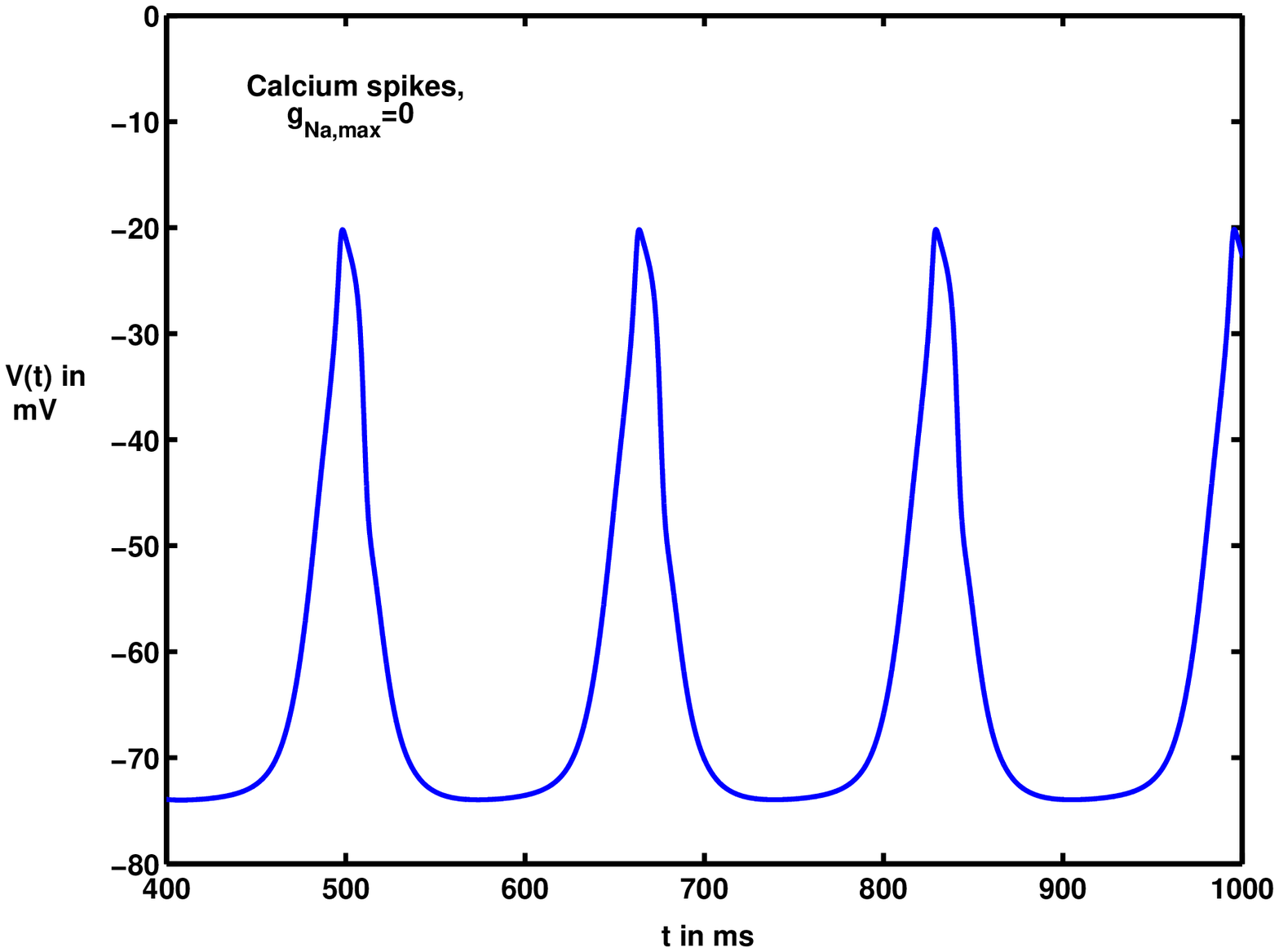,width=4.5 in}
\end{center}
\caption{Calcium spikes in the model without sodium current and
with several parameters modified.} 
\label{fig:wedge}
\end{figure}

The following changes were found to give rise to calcium spikes.
Since without the sodium current, the depolarizing effect of the T-type
current could not lead to sufficient opening of the high-threshold
N-type and/or L-type \CA channels as they were, the steady state activation and inactivation
functions of the N-type current,
\be m_{N,\infty}=\frac{1}{  1 + e^{  -(V -V_{N_1})/k_{N_1} }  }  \ee
 \be h_{N,\infty}=\frac{1}{  1 + e^{  (V -V_{N_3})/k_{N_3} }  }  \ee
were shifted by putting $V_{N_1}=-25$ rather than -10 mV and $V_{N_3}=-50$
rather than -45 mV. This would put the
N-type current in reach of the response to the T-type current. Note that 
the values  $V_{N_1}=-25$ and$V_{N_3}=-50$ had been employed
in producing regular spiking in Tuckwell and Penington (2012). 
To increase the N-type current, its maximal conductance 
$g_{N,max}$ was increased by 50\% to 0.06237 $\mu$S. 
The steady state acivation and inactivation functions of all 
other component currents were unchanged. However,
the maximal conductance of the T-type current was increased by 
a factor of 2 to $g_{T,max}$=0.4505 $\mu$S whereas the A-type 
potassium conductance was decreased by a factor of 4 to 0.1875 $\mu$S.
Two parameters controlling calcium dynamics required significant
change. Firstly, the calcium source factor CSF was reduced to
0.135 from 0.7 and finally, an extremely large adjustment
was required in the calcium pump strength $K_s$ by multiplying it by 30
to a value of 0.000011719 mM per ms. 
With these changes the calcium spikes depicted in Figure 6 were obtained
with an ISI of about 170 ms and an amplitude of about 54 mV.

\section{Discussion}
In the first part of this article we have summarized many biophysical
and physiological properties of DRN SE neurons, taken from various
electrophysiological and morphological studies performed in the last
40 or so years. Such a summary is expected to be useful
in the construction of computational models for these cells.
These models, when they include the many
neurotransmitters and neuromodulators which influence these cells,
including glutamate, GABA, corticotropin-releasing factor, orexin, substance P and
norepinephrine, will play an impoprtant role in understanding the 
key role of serotonergic neuron activity in controlling many 
known behavioral, physiological, cognitive and psychiatric phenomena. 
Some important circuits are the reciprocal connection between the DRN
several other brain structures including the
locus coeruleus, the hippocampus, the hypothalamus and the prefrontal cortex.
For reviews see the references given in the Introduction.

Secondly we have used a computational model
of these cells to see if it could generate, as observed experimentally,
calcium spikes when sodium current is blocked by TTX.
In the original model (Tuckwell and Penington, 2012) there are over
90 parameters, many of which were estimated from voltage-clamp data.
It has been demonstrated that with changes to the properties of N-type 
calcium current, which made it more readily excitable and with a somewhat
larger conductance, calcium spikes were obtained similar to
those obtained experimentally.  The T-type current was also
augmented and the opposing A-type potassium current was decreased.
There was a fine-tuning effect for the A- and T-type conductances.
If the A-type conductance was too small, the amplitude of the
spikes increased as well as the maxima of the internal \CA concentration. 
Increasing the T-type conductance countered this effect,
If the A-type conductance was too large, then internal \CA
decreased along with spike amplitude. The calcium pump strength $K_s$
and the value of CSF in (8) were the two parameters that had to be changed
considerably in order to obtain stable calcium spiking. If the calcium pump
was too weak then the ISI increased with time and \CA concentration
increased to non-physiological values. If $K_s$ were too large, then
the spike amplitude decreased steadily. If CSF was too large (above 0.135)
then the internal calcium concentration steadily increased. 
The reason for the necessity of such large changes in the parameters
$K_s$ and CSF needed to elicit a regular train of calcium spikes
is probably anatomical. The model is for a single compartment and
it is likely ´that calcium spikes are generated in the dendrites which
would result in quite different calcium dynamics, although
the experimental observations are made at the soma.
The large strength of the calcium pump is in accordance
with the requirement of a relatively small time constant of decay of internal 
calcium needed to generate calcium spikes in a model
of hippocampal neurons (Traub et al., 1991) where such
spikes were in fact dendritic in origin.

%

\u

\section{Acknowledgement} This article is dedicated to the memory
of Professor Luigi M. Ricciardi, advisor and friend. I also thank the
Max Planck Institute and Professor Dr Juergen Jost for support and 
Professor Laura Sacerdote of the University of Torino for her role in the organization of the BIOCOMP 2012 
meeting in Vietri and this volume.

\end{document}